\documentclass[%
 aip,
 amsmath,amssymb,
 reprint,%
]{revtex4-1}

\usepackage{graphicx}% Include figure files
\usepackage{dcolumn}% Align table columns on decimal point
\usepackage{bm}% bold math
\usepackage[utf8]{inputenc}
\usepackage[T1]{fontenc}
\usepackage{mathptmx}
\usepackage{etoolbox}
\usepackage{graphicx}% Include figure files
\usepackage{bm}% bold math
\usepackage{gensymb}
\usepackage{subcaption}
\usepackage{comment}
\usepackage{xcolor}
\usepackage{soul}

\def\d{\mathrm{d}}

\begin{document}

\preprint{AIP/123-QED}

\title[Airborne lifetime of respiratory droplets]{Airborne lifetime of respiratory droplets}
% Force line breaks with \\
\author{Avshalom Offner}
\email{avshalom.offner@ed.ac.uk}
 %\affiliation{School of Mathematics and Maxwell Institute for Mathematical Sciences, The University of Edinburgh, Edinburgh, UK}%Lines break automatically or can be forced with \\
\author{Jacques Vanneste}%
\affiliation{ 
School of Mathematics and Maxwell Institute for Mathematical Sciences, The University of Edinburgh, Edinburgh, UK}

\date{\today}% It is always \today, today,

\begin{abstract}
We formulate a model for the dynamics of respiratory droplets and use it to study their airborne lifetime in turbulent air representative of indoor settings. This lifetime is a common metric to assess the risk of respiratory transmission of infectious diseases, with longer lifetime correlating with higher risk. We consider a simple momentum balance to calculate the droplets spread, accounting for their size evolution as they undergo vaporization via mass and energy balances. The model shows how an increase in relative humidity leads to higher droplet settling velocity, which shortens the lifetime of droplets and can therefore reduce the risk of transmission. Emulating indoor air turbulence using a stochastic process, we numerically calculate probability distributions for the lifetime of droplets, showing how an increase in the air turbulent velocity significantly enhances the range of lifetimes. The distributions reveal non-negligible probabilities for very long lifetimes, which potentially increase the risk of transmission.
\end{abstract}

\maketitle

\section{Introduction}

Various viral diseases, including SARS-CoV-2, spread through respiratory transmission \cite{Bourouiba2021}. A leading precautionary measure to mitigate respiratory transmission is maintaining a `safe distance', reflecting \textit{a priori} knowledge on the spread of droplets and their time airborne. This distance, typically 2 meters, was calculated based on the model by Wells \cite{Wells1934}, who employed Stokes' solution to set the droplet velocity as proportional to its surface area, which decreases at a constant rate due to vaporization. This model, however, is  too simplistic to represent the dynamics of respiratory droplets, as was demonstrated by various recent works (e.g., Chong \textit{et al.} \cite{Chong2021} and Wang \textit{et al.} \cite{Wang2021a}). These works along with others \cite{Abuhegazy2020,Khosronejad2021,Olivieri2022} conducted direct numerical simulations on the dynamics of droplets in expiratory events, showing how small droplets can travel long distances and remain airborne longer than was considered previously. In the present work we formulate a particularly simple model to study the airborne lifetime of respiratory droplets, which allows us to simulate a large number of droplets and thus quantify the probability of anomalously long droplet airborne lifetime, which can have a disproportionate effect on virus transmission.

Droplets produced by expiratory events vaporize according to the vapor-pressure balance with the air that surrounds them, possibly after a short period of growth \cite{Ng2021}. Saliva droplets do not completely vaporize due to non-volatile substances, resulting in droplets saturating to a small finite size as they reach equilibrium with the surrounding air \cite{Liu2017,Dhand2020}. Here, we formulate a model for the vaporization of saliva droplets to capture this unique size evolution, and calibrate it with experimental measurements \cite{Lieber2021}. This model is then coupled with a simple momentum balance to describe the dynamics of a single, representative droplet. To retain our model simplicity, we consider a constant relative humidity in the air, thus not accounting for the humid turbulent puff exhaled with the droplets \cite{Bourouiba2014}. This prevents our analysis from describing the dynamics of small droplets (with initial radius $R_{0}\lesssim10\,\mathrm{\mu m}$), which initially travel within a puff of high humidity that slows their vaporization and increases their airborne lifetime \cite{Chong2021,Wang2021a,Bourouiba2020}.

\section{\label{sec:level1}Model}

Droplets expelled from the human body span a wide range of sizes, from less than a micron to hundreds of microns in diameter \cite{Papineni1997,Yang2007a}. The vast majority of expelled droplets travel at sufficiently low velocity \cite{Bahl2021} to satisfy $Re=\left|\bm{v}-\bm{u}\right|R/\nu_{a}\ll1$, where $Re$ is the Reynolds number, $\bm{v}$ and $\bm{u}$ the droplet and air velocities (\textbf{bold} letters denote vector quantities), respectively, $R$ the droplet radius, and $\nu_{a}$ the air kinematic viscosity. Taking advantage of the small Reynolds number, we describe the dynamics of airborne droplets using the aerosol limit of the Maxey--Riley model \cite{Maxey1987}
\begin{align}
\frac{\d\bm{x}}{\d t}&=\bm{v}, \\
m\frac{\d\bm{v}}{\d t}&=-\frac{6\pi R\mu_{a}}{C} \left(\bm{v}-\bm{u}\right)-m\bm{g}, \label{eq:dv dimensional}
\end{align}
where $\bm{x}$ is the droplet position, $m=4\pi R^{3}\rho/3$ is its mass, with $\rho$ the droplet density, $\mu_{a}$ is the air dynamic viscosity, $\bm{g}$ is the gravitational acceleration, and $C$ is the slip correction for drag force over small spheres \cite{Knudsen1911}
\begin{equation}
C=1+\frac{l}{R}\left[\mathcal{A}+\mathcal{B}\exp\left(-\mathcal{C}\frac{R}{l}\right)\right], \label{eq:slip correction dimensional}
\end{equation}
where $l$ is the air mean free path, and $\mathcal{A}$, $\mathcal{B}$ and $\mathcal{C}$ are constants determined by fitting (\ref{eq:slip correction dimensional}) to experimental results. In this work we use the values calculated by Davies \cite{Davies1945} (see table \ref{tab:parameters}). Eq.\ (\ref{eq:dv dimensional}) is obtained from the complete Maxey--Riley model \cite{Maxey1983} by neglecting all terms proportional to the air-to-liquid density ratio
$\vartheta_{\rho}=\rho_{a}/\rho\ll1$. Thus we treat droplets as liquid particles, small enough to be considered rigid (and therefore spherical) and heavy enough that $\vartheta_{\rho}\approx10^{-3}\ll1$.

Volatile droplets lose mass while vaporizing and hence their radius varies. To account for this variation, we include a mass balance equation which depends on the droplet temperature. This temperature is in turn governed by energy balance, leading to the system
\begin{align}
\frac{\d m}{\d t}&=-\frac{\overline{h}_{m}A M_{w}}{\mathfrak{R}}\left(\frac{p}{T}-\frac{p_{a}}{T_{a}}\right), \label{eq:dS dimensional} \\
m c_{p}\frac{\d T}{\d t}&= H\frac{\d m}{\d t}-\overline{h}_{h}A\left(T-T_{a}\right), \label{eq:dT dimensional}
\end{align}
where $A=4\pi R^{2}$ is the droplet surface area, $p$ and $T$ are the droplet vapor pressure and temperature, $p_{a}$ and $T_{a}$ their counterparts for air, $\mathfrak{R}$ is the universal gas constant, $M_{w}$ is water's molecular weight, $c_{p}$ is the droplet isobaric heat capacity, $H$ is the heat of vaporization, and $\overline{h}_{m}$ and $\overline{h}_{h}$ are the convective mass and heat transfer coefficients, respectively \cite{Incropera2010,Ranz1952}. In (\ref{eq:dS dimensional})--(\ref{eq:dT dimensional}) we assume that convection dominates over diffusion between the droplet and air. Next, we non-dimensionalize variables by letting
\begin{align}
t&=\tau\hat{t}\,,\,\bm{v}=U\hat{\bm{v}}\,,\,\bm{u}=U\hat{\bm{u}},\, \,  \bm{x}=\tau U\hat{\bm{x}}, \nonumber \\
R&=R_{0}\hat{R}\,,\,\bm{g}=\frac{U}{\tau}\hat{\bm{g}}\,,\,T=T_{a}\hat{T}\ \textrm{and} \ p=P\hat{p}, \label{eq:scaling}
\end{align}
where the hat denotes dimensionless quantities, $U$ is a characteristic velocity, $R_{0}\equiv R\left(t=0\right)$ is the droplet initial radius, $\tau=2R_{0}^{2}/\left(9\nu_{a}\vartheta_{\rho}\right)$ is the Stokes time scale, and $P$ is the atmospheric pressure. To simplify our model, we assume that $Pr=Sc=\lambda$, where $Pr=\mu_{a}c_{p,a}/k_{a}$ is the \textit{Prandtl} number with $c_{p,a}$ and $k_{a}$ the air isobaric heat capacity and thermal conductivity, and $Sc=\nu_{a}/D$ is the \textit{Schmidt} number with $D$ the molecular diffusion coefficient for air-water vapor mixture. This assumption is not strictly necessary for the numerical calculations that follow, however it greatly assists in drawing physical insight from the model with only a marginal sacrifice in accuracy (see table \ref{tab:parameters} for typical values of $Pr$ and $Sc$). Building upon the scaling theory for convection over a sphere \cite{Incropera2010}, the mass and heat convection dimensionless parameters -- the \textit{Sherwood} and \textit{Nusselt} numbers, $Sh\left(Re,Sc\right)=\overline{h}_{m}R/D$ and $Nu\left(Re,Pr\right)=\overline{h}_{h}R/k_{a}$, respectively -- are identical for $Pr=Sc$, and therefore we write $Sh=Nu=\mathcal{F}$. Finally, we write the model in dimensionless form, omitting all hats for convenience and recalling that $\d m=\rho A\d R$ to obtain
\begin{subequations}\label{eq:ODE system}
\begin{align}
\frac{\d\bm{x}}{\d t}&=\bm{v}, \label{eq:dx ND} \\
\frac{\d\bm{v}}{\d t}&=-\left(\frac{\bm{v}-\bm{u}}{C\,S}+\bm{g}\right), \label{eq:dv ND} \\
\frac{\d S}{\d t}&=-\frac{4\vartheta_{M}\Phi\left(S,\theta\right)}{9\lambda}\mathcal{F}\left(Re,\lambda\right), \label{eq:dS ND} \\
\frac{\d\theta}{\d t}&=-\frac{2\left[\mathcal{H}_{d}\vartheta_{M}\Phi\left(S,\theta\right)+\vartheta_{c}\theta\right]}{3\lambda\,S}\mathcal{F}\left(Re,\lambda\right), \label{eq:dtheta ND}
\end{align} 
\end{subequations}
where $S\equiv R^{2}$, $\theta\equiv T-1$ is the dimensionless temperature difference between the droplet and air, $\vartheta_{M}=M_{w}/M_{a}$ and $\vartheta_{c}=c_{p,a}/c_{p,d}$ are the ratios between water and air molecular mass and heat capacity, respectively, $\mathcal{H}_{d}=H/\left(c_{p} T_{a}\right)$ is the droplet latent heat parameter, $C\left(S;Kn\right)$ is the dimensionless form of (\ref{eq:slip correction dimensional}) with $Kn=l/R_{0}$ the \textit{Knudsen} number, and $\Phi$ is the scaled vapor pressure difference between the droplet and air. The latter is expressed through the Clausius-Clapeyron relation, in the form
\begin{align}
\Phi&=\exp\left[-\mathcal{H}_{v}\left(\frac{1}{\theta+1}-\frac{1}{T_{b}+\beta S^{-3/2}}\right)+\frac{\eta}{S^{1/2}}\right]/\left(\theta+1\right) \nonumber \\
&-\phi\exp\left[-\mathcal{H}_{v}\left(1-\frac{1}{T_{b}}\right)\right], \label{eq:Phi}
\end{align}
where $\phi$ is the air relative humidity, which is assumed to be constant, $\mathcal{H}_{v}=HM_{w}/\left(\mathfrak{R}\,T_{a}\right)$ is the vapor latent heat parameter, and $T_{b}$ is the water scaled boiling temperature. The term $\eta/S^{1/2}$ is the addition of capillary evaporation \cite{Kelvin1871}, stemming from the droplet curvature, where $\eta=2\gamma_{c}\vartheta_{\rho}\vartheta_{M}/\left(PR_{0}\right)$ is the scaled surface tension with $\gamma_{c}$ the water-air surface tension. The term $\beta S^{-3/2}$, with $\beta$ a constant, is the correction to the droplet boiling temperature due the increasing concentration of non-volatile substances in its composition \cite{Atkins1990}. Here we assume that saliva may be modelled as a dilute solution throughout its vaporization, such that the loss of mass from the droplet is expressed only through a change in volume -- which is proportional to $S^{3/2}$ -- while the droplet density remains that of water, independent of the (small) concentration of non volatiles. The dimensionless constant $\beta$ depends on the saliva reference composition; its value is determined below using experimental measurements of saliva droplet vaporization.

\begin{table}[]
    \centering
    \begin{tabular}{c|c|c}
    \textbf{Notation} & \textbf{Interpretation} & \textbf{Value @ 20\degree C} \\
    \hline
    $Pr$ & Air \textit{Prandtl} number & 0.7 \\
    $Sc$ & Air--water \textit{Schmidt} number & 0.62 \\
    $\lambda$ & Approximation for $Pr$ and $Sc$ & 0.66 \\
    $g$ & Scaled gravity & 0.35 \\
    $\vartheta_{\rho}$ & Air--water density ratio & $1.2\cdot10^{-3}$ \\
    $\vartheta_{c}$ & Air--water heat capacity ratio & 0.24 \\
    $\vartheta_{M}$ & Water--air molecular weight ratio & 0.62 \\
    $\mathcal{H}_{d}$ & $H/\left(c_{p}T_{a}\right)$ & 1.8 \\
    $\mathcal{H}_{v}$ & $HM_{w}/\left(\mathfrak{R}T_{a}\right)$ & 16 \\
    $T_{b}$ & Scaled boiling temperature & 1.27 \\
    $\eta$ & Scaled surface tension & 0.009 \\
    $\beta$ & Saliva compound constant & $8\cdot10^{-4}$ \\
    $\mathcal{F}$ & Convective mass/heat & 0.707 \\
    $\,$ & transfer coefficient & $\,$ \\
    $Kn$ & Droplet \textit{Knudsen} number & 0.002 \\
    $\mathcal{A,B,C}$ & Eq. (\ref{eq:slip correction dimensional}) coefficients \cite{Davies1945} & $1.257,0.4,1.1$
    \end{tabular}
    \caption{A list of all dimensionless parameters in our model, along with representative values calculated for $R_{0}=30\:\mathrm{\mu m}$ and $T=T_{a}=20\degree\mathrm{C}$. Reference values for properties of air and water were taken from Poling \textit{et al}. \cite{Poling2001}.}
    \label{tab:parameters}
\end{table}

To make the system (\ref{eq:dx ND})--(\ref{eq:dtheta ND}) complete, an explicit form must be assigned to $\mathcal{F}\left(Re,\lambda\right)$. In order to recover the well-known $D^2$ law of vaporization \cite{Spalding1950}, i.e. that $S$ decreases linearly with $t$, $\mathcal{F}$ must be constant. This decouples (\ref{eq:dS ND})--(\ref{eq:dtheta ND}) from the droplet dynamics (\ref{eq:dx ND})--(\ref{eq:dv ND}) and forms a closed system. To determine the value of $\mathcal{F}$, one could invoke Ranz and Marshall's theoretical result \cite{Ranz1952} giving $\mathcal{F}=2$ for convection over a single sphere in the limit $Re\rightarrow0$. Instead, we  fit values to $\mathcal{F}$ and $\beta$ according to experimental measurements of saliva droplet vaporization.

\subsection{Droplet vaporization}

\begin{figure}
    \centering
    \includegraphics[trim=0 0 1.3cm 1.2cm,clip,scale=0.4]{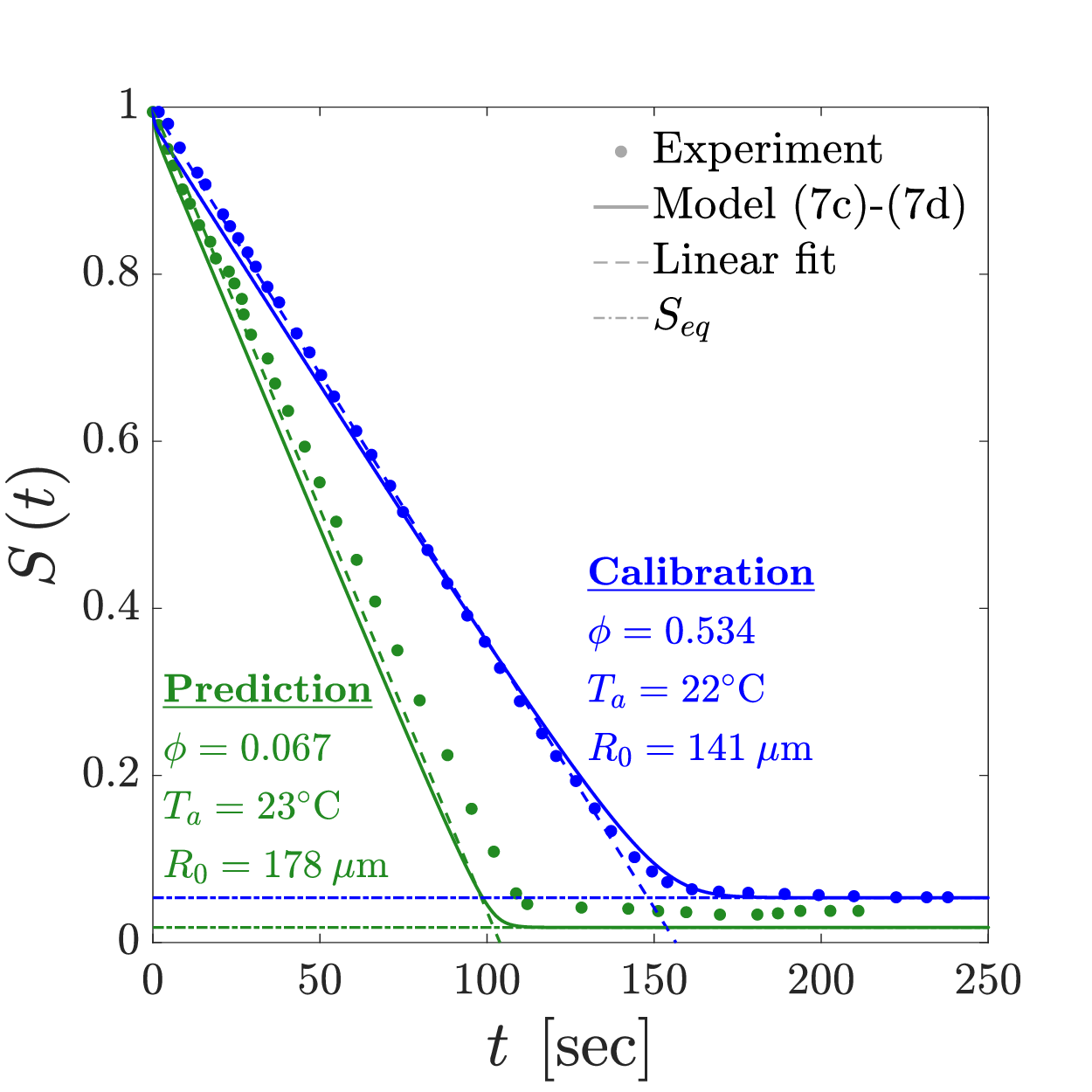}
    \caption{Time evolution of saliva droplets vaporizing in air, expressed as the dimensionless radius squared, $S\left(t\right)\equiv R\left(t\right)^{2}$. The dots mark experimental measurements from Lieber \textit{et al.} \cite{Lieber2021}, solid lines are the numerical solution to (\ref{eq:dS ND})-(\ref{eq:dtheta ND}), dashed and dashed dotted lines are the $D^2$ linear laws with slopes $\alpha$ and the equilibrium size $S_{eq}$, calculated through (\ref{eq:alpha}) and (\ref{eq:Seq}), respectively. The experimental results for $\phi=0.534$ (blue) were used to fit the model constants -- $\mathcal{F}=0.707$ and $\beta=8\cdot10^{-4}$ -- and hence theory and experiment match by construction. These values were then used to predict the time evolution at $\phi=0.067$ (green), showing good quantitative agreement. The results indicate that the droplet size evolution may be approximated as a piecewise function with a linear decay followed by an equilibrium value, $S_{eq}$.}
    \label{fig:S(t) comparison}
\end{figure}

The vaporization process of saliva droplets may be divided into three distinct stages: initial cooling, $D^2$ vaporization, and saturation to equilibrium. Figure \ref{fig:S(t) comparison} shows the size evolution of saliva droplets vaporizing in air, where the two last stages are clearly depicted -- the linear decrease marking the $D^2$ vaporization stage and the saturation to a constant value corresponding to the last stage. The initial cooling stage typically lasts less than a second and is therefore difficult to observe in figure \ref{fig:S(t) comparison}. Below we describe the process in each of these stages, and derive analytic approximations that are key for the analysis that follows. 

\subsubsection{Stage 1: initial cooling}

Droplets expelled from a human body are typically warmer than the surrounding air, i.e. $\theta\left(t=0\right)>0$. The vaporization process begins with a rapid decrease in $\theta$ as both vaporization and convection -- $\mathcal{H}_{d}\vartheta_{M}\Phi$ and $\vartheta_{c}\theta$ in (\ref{eq:dtheta ND}), respectively -- cool the droplet, while its size remains nearly unchanged. As $\theta$ falls below zero, convection heats the droplet until a balance with vaporization is reached and its temperature stabilizes. The temperature then remains nearly constant throughout the second vaporization stage, and therefore we denote it by $\theta_{D2}$. The value of $\theta_{D2}$ may be computed numerically by setting the right-hand-side (RHS) of (\ref{eq:dtheta ND}) to zero and neglecting the terms $\beta S^{-3/2}$ and $\eta S^{-1/2}$ in (\ref{eq:Phi}), since these only become significant for $S\ll1$, whereas here $S\approx1$. 

\subsubsection{Stage 2: $D^2$ vaporization}

Once a droplet temperature stabilizes at $\theta_{D2}$, the RHS of (\ref{eq:dS ND}) becomes nearly constant, giving rise to a linear decrease of $S$ with time -- the well-known $D^2$ law \cite{Spalding1950}. The slope,  
\begin{equation}
\alpha=-\frac{4\vartheta_{M}\mathcal{F}\Phi\left(\theta_{D2}\right)}{9\lambda}, \label{eq:alpha}
\end{equation}
can be estimated by fitting a straight line to  experimental measurements of $S$ vs.\ $t$ in the vaporizing stage. \ \eqref{eq:alpha} then yields an estimate for the value of the constant $\mathcal{F}$. We use the results of Lieber \textit{et al.} \cite{Lieber2021}, who recorded the vaporization of levitating saliva droplets, to evaluate $\mathcal{F}$ by least-squares fitting to the data with $\phi=0.534$ in figure \ref{fig:S(t) comparison} (blue), yielding $\mathcal{F}=0.707$. This value is then substituted back to (\ref{eq:dS ND})-(\ref{eq:dtheta ND}) and used to predict the vaporization at $\phi=0.067$ (green), showing a good quantitative agreement. Encouraged by this agreement, we use the model to describe the vaporization throughout the calculations that follow. 

\subsubsection{Stage 3: saturation to equilibrium}

\begin{figure}
    \centering
    \includegraphics[trim=0 1cm 0 2cm,clip,scale=0.36]{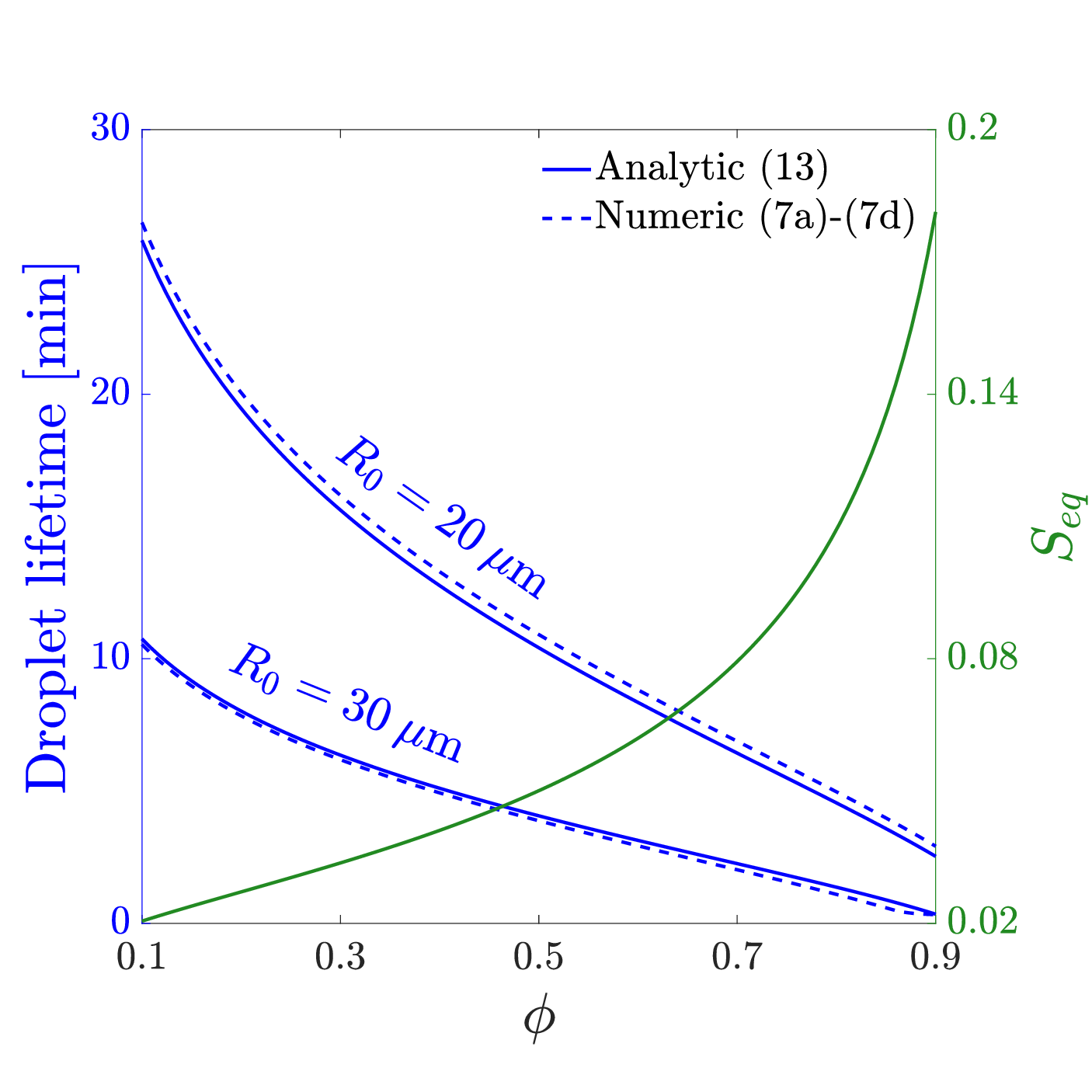}
    \caption{\textbf{Left}: airborne lifetime of droplets released $1.5\,\mathrm{m}$ above the ground in quiescent air for two initial radii, $R_{0}=20$ and $30\:\mathrm{\mu m}$, as a function of relative humidity, $\phi$. The solid curves are calculated through (\ref{eq:tf analytic}) and the dashed curves are obtained by solving (\ref{eq:dx ND})-(\ref{eq:dtheta ND}) numerically with $u=0$. The analytic and numeric results are in very good agreement, ratifying that (\ref{eq:S piecewise}) is a good quantitative approximation for $S\left(t\right)$. The monotonically decreasing trend of the curves emphasizes that an increase in $\phi$ decreases the lifetime of respiratory saliva droplets. \textbf{Right}: droplet equilibrium size expressed as the scaled radius squared, $S_{eq}$, as a function of $\phi$. An increase in $\phi$ sets the equilibrium between the saliva and air at a lower concentration of non-volatiles, which increases the droplet equilibrium size.}
    \label{fig:Analytic results}
\end{figure}

As water vaporizes from a droplet and its size diminishes, the concentration of non-volatiles increases, leading to an increase in the droplet boiling temperature. This process is accounted for by the term $\beta S^{-3/2}$ in (\ref{eq:Phi}), recalling that $S^{3/2}$ is proportional to the droplet volume. Following the assumption of constant relative humidity, the droplet-air vapor pressure gradient decreases with a decrease in $S$ until, at a finite size $S_{eq}>0$, the droplet and air reach a state of thermodynamic equilibrium and the vaporization terminates. This equilibrium corresponds to the vanishing of the right-hand side of (\ref{eq:dS ND}) and (\ref{eq:dtheta ND}), resulting in $\theta=\Phi=0$. This readily provides the means for calculating $S_{eq}$, by solving $\Phi\left(S_{eq},\theta_{eq}=0\right)=0$. The limit $\eta \to 0$, which neglects the minor effect of surface curvature, gives the simple expression
\begin{equation}
S_{eq}=\left[\frac{-\beta\left(\mathcal{H}_{v}+T_{b}\log\phi\right)}{T_{b}^{2}\log\phi}\right]^{2/3}. \label{eq:Seq}
\end{equation}
By comparing (\ref{eq:Seq}) to the droplet equilibrium size measured experimentally, we fit the constant $\beta=8\cdot10^{-4}$. The dependence of $S_{eq}$ on temperature is weak as both $\mathcal{H}_{v},T_{b}\propto T_{a}^{-1}$, yielding $S_{eq}\propto T_{a}^{2/3}$, which varies very little in the indoor temperature range $285$--$305\;\mathrm{K}$. Relative humidity, on the hand, strongly affects the equilibrium size, with $S_{eq}$ increasing with $\phi$ as shown in figure \ref{fig:Analytic results} (green curve and right vertical axis). As a result, droplets at high relative humidity are larger and therefore fall faster to the ground.

The results in figure \ref{fig:S(t) comparison} demonstrate that a saliva droplet size evolution may be approximated by the piecewise form
\begin{equation}
S\left(t\right)\approx\begin{cases}
1-\left|\alpha\right| t, & t<t_{eq}\\
S_{eq}, & t>t_{eq}
\end{cases}, \label{eq:S piecewise}
\end{equation}
in which $t_{eq}=\left(1-S_{eq}\right)/\left|\alpha\right|$, with only a marginal sacrifice in accuracy. This approximation is used to obtain analytic results for the droplet airborne lifetime in the next section.

\subsection{Droplet dynamics}

We now turn our focus to  the dynamics of respiratory saliva droplets, governed by (\ref{eq:dx ND})--(\ref{eq:dv ND}) with $S\left(t\right)$ given by the solution to (\ref{eq:dS ND})--(\ref{eq:dtheta ND}). We concentrate our analysis on the vertical motion of droplets as it determines the lifetime -- the time for a droplet released from a height $z_{0}>0$ to reach the ground  -- which is a key metric for assessing the risk of respiratory transmission. We note that care needs to be exercised in inferring viral transmission risks from droplet airborne lifetimes because aerosolized viruses can be deactivated at varying environmental conditions. Herinafter, $z,\, v,\, u$ and $g$ are scalar quantities representing vertical components, i.e. aligned with gravity. We begin our analysis by considering the simplest case of $u=0$, corresponding to a room of completely quiescent air.

\subsubsection{Quiescent air}

The case $u=0$ provides a benchmark result for free-falling droplets, from which valuable physical insight can be drawn. By employing the piecewise form (\ref{eq:S piecewise}) of $S$, we derive an analytic solution for $z\left(t\right)$. The explicit solution to $z\left(t\right)$ is tedious and difficult to interpret (see appendix \ref{ap:A}). However, using the smallness of $Kn$ for droplets larger than $R_{0}\gtrsim 5\,\mathrm{\mu m}$ (see table \ref{tab:parameters})  we simplify this expression by focusing on the motion after the vaporization terminates and taking a series expansion about $Kn=0$, yielding
\begin{align}
z\left(t\right)&\approx z_{0}+v_{0}\left(1+\mathcal{A}Kn\right)-\frac{g}{6\left|\alpha\right|}\left(3+4\mathcal{A}Kn\right) \nonumber \\
&-g\left(S_{eq}+\mathcal{A}Kn\sqrt{S_{eq}}\right)\left(t-t_{eq}\right),\quad t>t_{eq},\label{eq:x(t) analytic}
\end{align}
with $z_{0}$ and $v_{0}$ the droplet initial height and vertical velocity, respectively. In deriving (\ref{eq:x(t) analytic}) we considered the case $\left|\alpha\right|\ll1$ and $S_{eq}^{3/2}\ll1$, which holds throughout the realisable range for $T_{a}$ and $\phi$ (see appendix \ref{ap:A} for the complete derivation). The first and second terms on the RHS of (\ref{eq:x(t) analytic}), $z_{0}+v_{0}\left(1+\mathcal{A}Kn\right)$, denote the droplet initial conditions, where $\left|v_{0}\right|\ll z_{0}$ is typically obtained since $z\propto\tau^{-1}$ in our scaling (\ref{eq:scaling}). For simplicity, all the results for quiescent air are calculated with $v_{0}=0$. The third term represents the altitude decrease by time $t_{eq}$, inversely proportional to the $D^2$ vaporization rate, $\left|\alpha\right|$, which decreases nearly linearly with $\phi$ (see appendix \ref{ap:B}). This indicates that an increase in relative humidity increases the altitude drop during the vaporization stage, which shortens the droplet airborne lifetime. The fourth term recovers the expected linear descent at terminal velocity $v_{t}=-g\left(S_{eq}+\mathcal{A}Kn\sqrt{S_{eq}}\right)$, involving $S_{eq}$ that increases with an increase in $\phi$. This demonstrates how an increase in $\phi$ translates to higher droplet settling velocities that lead to shorter droplet airborne lifetimes. All the terms proportional to $Kn$, which stem from velocity slip on the droplet surface, naturally increase the droplet settling velocity since the drag force is reduced. By setting $z\left(t\right)=0$ in (\ref{eq:x(t) analytic}) and substituting $t_{eq}=\left(1-S_{eq}\right)/\left|\alpha\right|$, one easily derives an analytic approximation for a droplet airborne lifetime in quiescent air with $v_{0}=0$,
\begin{equation}
t_{f}\approx\frac{z_{0}\left(1-\mathcal{A}Kn\,S_{eq}^{-1/2}\right)}{g\,S_{eq}}-\frac{3\left(1-2S_{eq}\right)-\mathcal{A}Kn\left(4-3S_{eq}^{-1/2}\right)}{6\alpha\,S_{eq}}. \label{eq:tf analytic}
\end{equation}
Figure \ref{fig:Analytic results} (blue curve, left vertical axis) shows the lifetime of two respiratory droplets ($R_{0}=20,30\:\mathrm{\mu m}$) in quiescent air as a function of relative humidity, calculated both analytically through (\ref{eq:tf analytic}) as well as numerically by solving (\ref{eq:dx ND})-(\ref{eq:dtheta ND}) with $u=0$ (solid and dashed curves, respectively). The curves clearly show that (i) a droplet airborne lifetime monotonically decreases with an increase in $\phi$ and that (ii) the analytic approximation (\ref{eq:tf analytic}) closely follows the numerical result, verifying the use of (\ref{eq:S piecewise}) to approximate $S\left(t\right)$ as well as the asymptotic approximations $\left|\alpha\right|\ll1$ and $S_{eq}^2\ll1$. 

\subsubsection{Indoor turbulence}

The air motion in indoor settings is turbulent, driven by natural and forced ventilation, people motion and other factors. To retain the model simplicity, we represent indoor turbulence by an Ornstein--Uhlenbeck process
\begin{equation}
\d u=-\gamma u\d t+\sigma\d W, \label{eq:O-U}
\end{equation}
in which $W$ is a Wiener process, $\gamma=0.58\:\mathrm{s^{-1}}$ is fitted using experimental measurements of indoor air velocity \cite{Loomans1998}, and $\sigma=\sqrt{2\gamma}U$ with $U$ the root-mean-square (RMS) velocity. Any effect of mean air velocity on the droplet motion can be incorporated in future work by accounting for air flow in specific settings and conditions.

\section{Results and discussion}

\begin{figure}
\centering
\subcaptionbox{\label{fig:U01}}
{\includegraphics[trim=0.8cm 0 0.6cm 1.2cm,clip,scale=0.26]{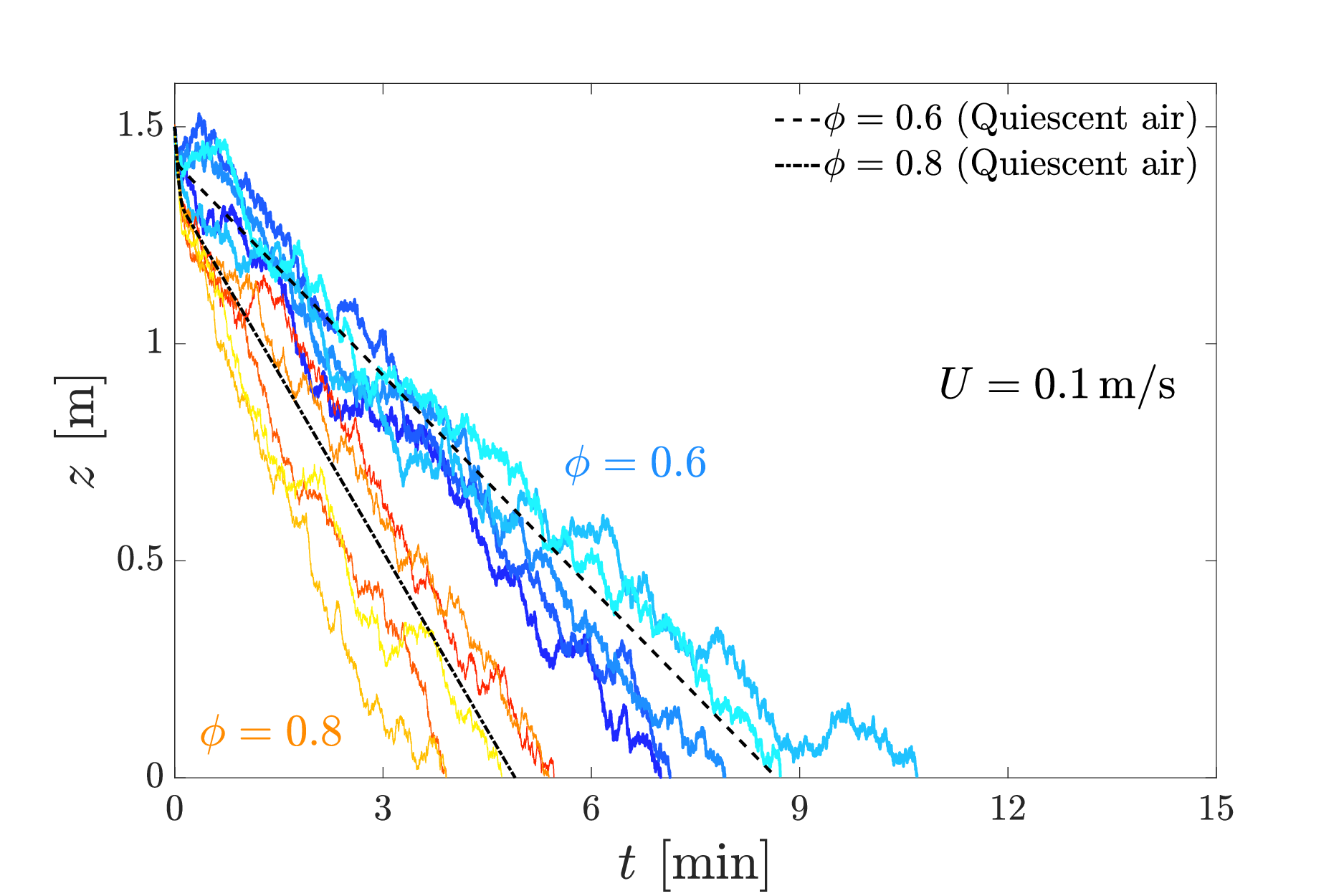}}
\subcaptionbox{\label{fig:U03}}
{\includegraphics[trim=0.8cm 0 0.6cm 1.2cm,clip,scale=0.26]{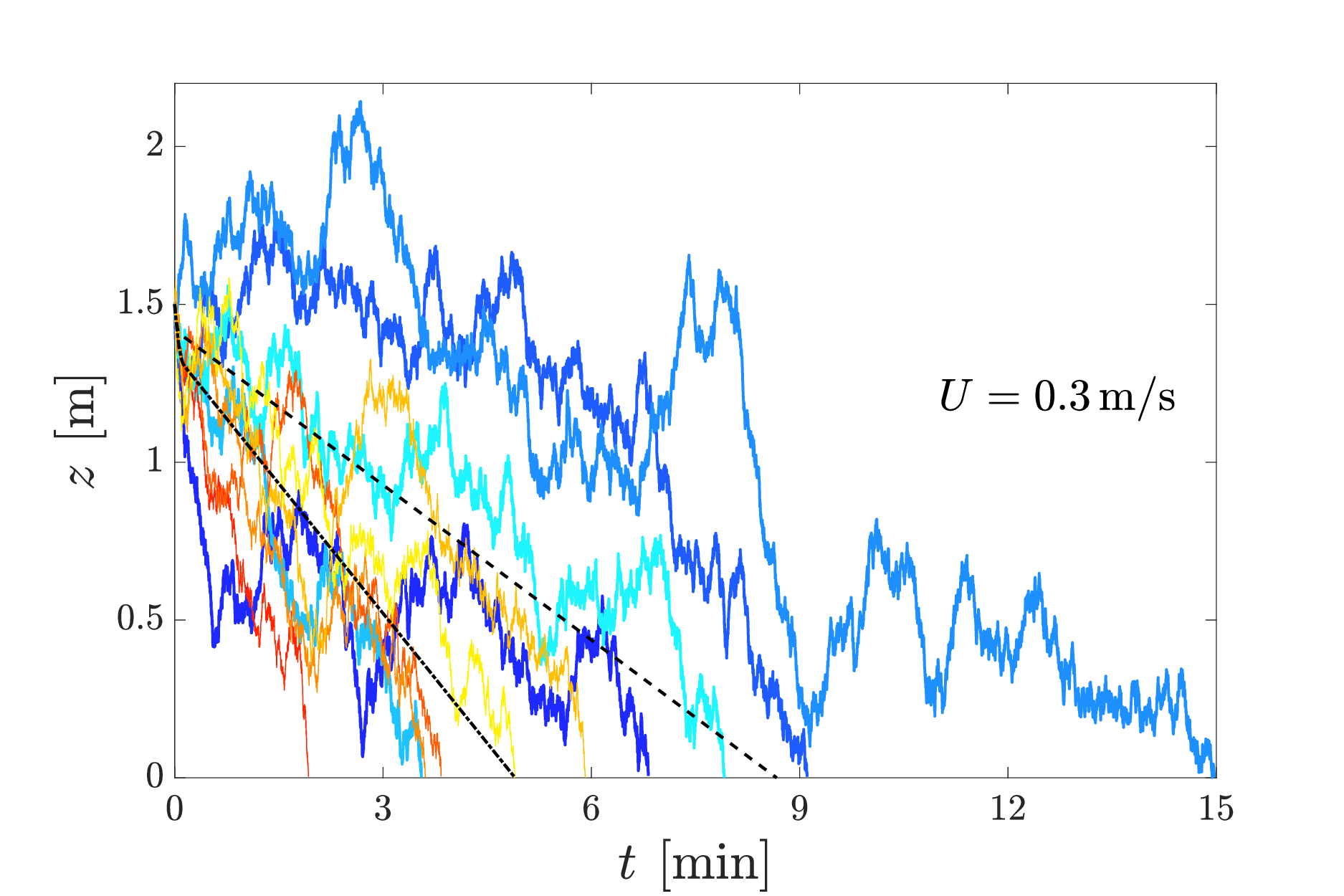}}
\caption{Representative trajectories of the vertical motion of saliva droplets at relative humidities $\phi=0.6$ (cyan to dark blue) and $\phi=0.8$ (yellow to red), for air root-mean-square velocity (a) $U=0.1$ and (b) $U=0.3\:\mathrm{m/s}$. The black dashed ($\phi=0.6$) and dashed-dotted ($\phi=0.8$) curves are the deterministic trajectories for a droplet in quiescent air, calculated from (\ref{eq:x(t) analytic}). }\label{fig:Representative results}
\end{figure}

We calculate the airborne lifetime of droplets, defined as the time for a droplet released at $z=1.5\;\mathrm{m}$ to reach the ground, by solving (\ref{eq:dx ND})--(\ref{eq:dtheta ND}) and (\ref{eq:O-U}) numerically. In what follows, we only present results for droplets in the range $R_0=20-50\,\mathrm{\mu m}$; calculation with smaller droplets replicated the qualitative behavior of $R_0=20\,\mathrm{\mu m}$, and larger droplets reach the ground within seconds regardless of the relative humidity and air velocity. Figure \ref{fig:Representative results} show trajectories of 5 droplets with $R_{0}=20\:\mathrm{\mu m}$ at relative humidity $\phi=0.6$ (cyan to dark blue colors) and $0.8$ (yellow to red). The random trajectories obtained in turbulent air with RMS velocities (a) $U=0.1$ and (b) $0.3\:\mathrm{m/s}$ are compared with the deterministic trajectories obtained for quiescent air (black dashed and dashed-dotted curves). The distinction between the $\phi=0.6,0.8$ bundles of trajectories in figure \ref{fig:U01} demonstrates that the decrease in droplet airborne lifetime as $\phi$ is increased, predicted analytically for quiescent air, also holds for $u\neq0$. An increase in air velocity results in greater variability of droplet airborne lifetime -- as clearly seen in figure \ref{fig:U03} -- which stems from the enhanced entrainment of droplets with the turbulent air flow.

\begin{figure}
    \centering
    \subcaptionbox{\label{fig:pdfU01}}
    {\includegraphics[trim=4.0cm 0 0 1cm,clip,scale=0.22]{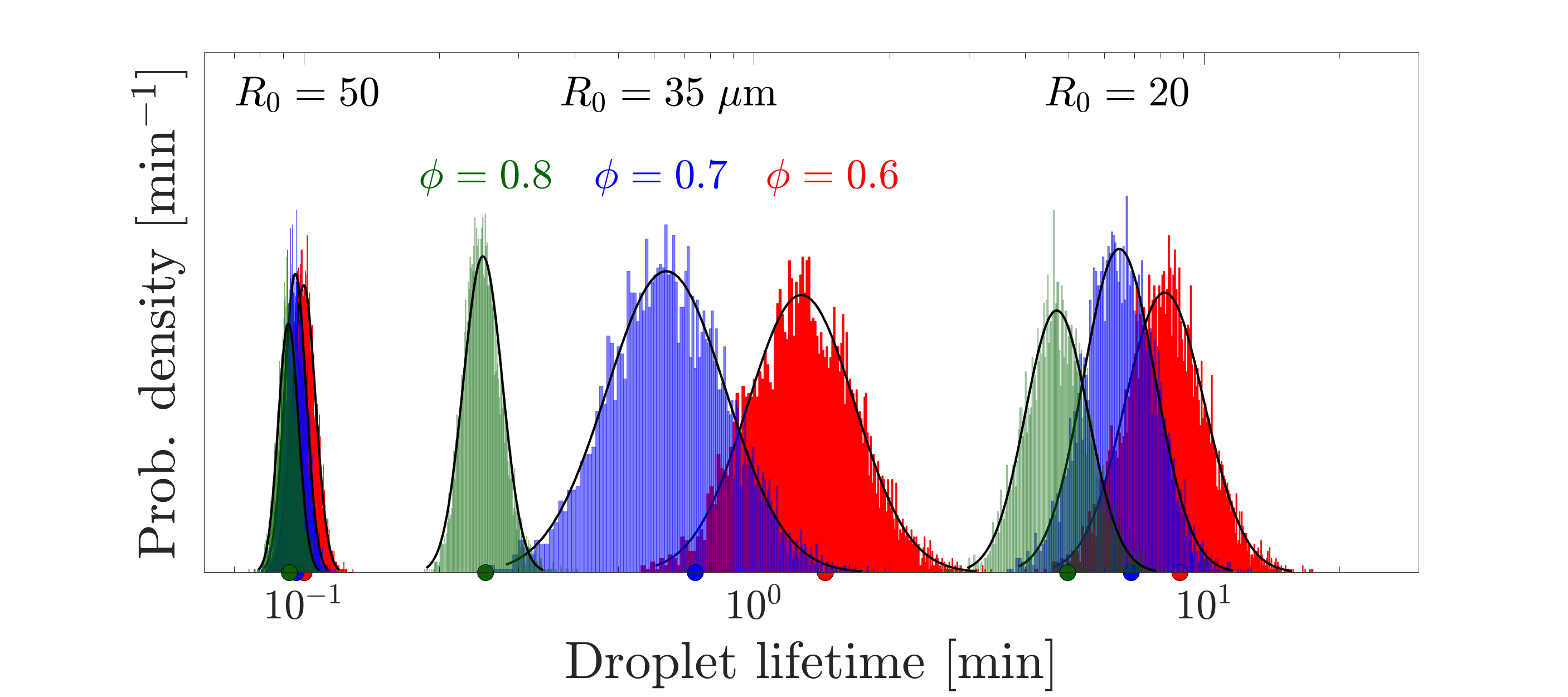}}
    \subcaptionbox{\label{fig:pdfU03}}
    {\includegraphics[trim=4.0cm 0 0 0,clip,scale=0.22]{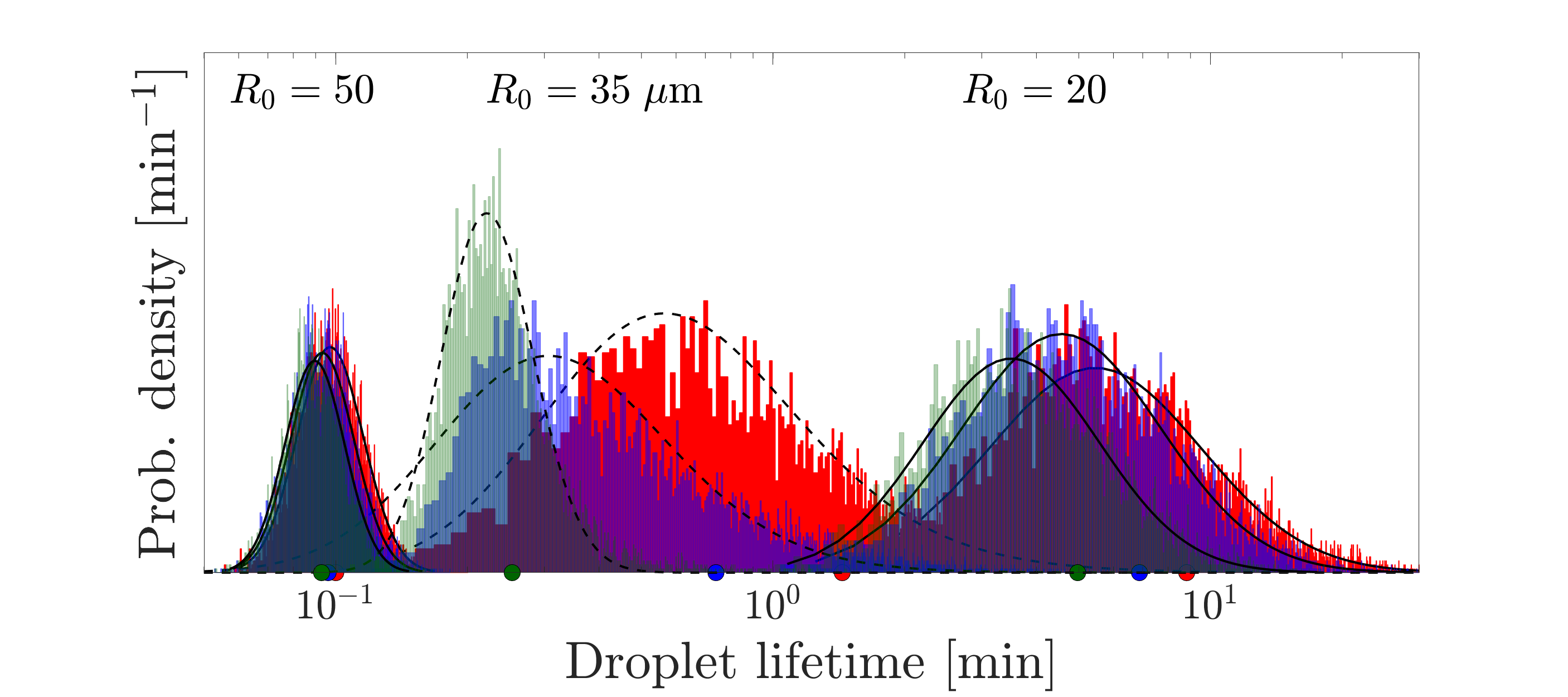}}
    \caption{Probability density function for the airborne lifetime of respiratory droplets released $1.5\:\mathrm{m}$ above ground with initial radii of $20$ (right), $35$ (center), and $50\;\mathrm{\mu m}$ (left). The red, blue, and green histograms correspond to relative humidity of $\phi=0.6,\,0.7$ and $0.8$, respectively. Solid and dashed black curves are lognormal and log-lognormal distributions fitted to the data, respectively. The air root-mean-square velocity is (a) $U=0.1$ and (b) $U=0.3\;\mathrm{m/s}$. Markers on the horizontal, logarithmic axis mark the analytic result for quiescent air (colors matching the histograms). As expected, smaller droplets remain airborne longer. An increase in relative humidity slows the vaporization rate and increases droplets' equilibrium size, thus accelerating their descent to the ground and shortening their lifetime. An increase in air velocity increases the lifetime variability.}
    \label{fig:PDFs}
\end{figure}

The findings above can clearly be noted by observing random droplet trajectories, such as the ones depicted in figure \ref{fig:Representative results}. In the context of disease transmission, however, it is essential to quantify the probability for anomalously long lifetimes which can dramatically increase the rate of transmission. Accordingly, we characterize the entire range of lifetimes statistically by calculating 5,000 droplet trajectories and collating their lifetimes into a probability density function (PDF). Figure \ref{fig:PDFs} show these PDFs for saliva droplets with initial radii $R_{0}=20,\, 35$ and $50\;\mathrm{\mu m}$, at relative humidity of 60\% (red), 70\% (blue), and 80\% (green), for (a) $U=0.1$ and (b) $U=0.3\:\mathrm{m/s}$. The ambient indoor temperature is $20\degree\mathrm{C}$. We emphasize that the horizontal axis is logarithmic, demonstrating the extensive variability in droplet airborne lifetime. The black solid and dashed curves are distributions fitted to the data (discussed below), and the markers on the horizontal axis give the lifetimes for $u=0$ (colors matching the histograms), which nearly coincide with the mean values of the PDFs throughout our calculations. This agreement was anticipated, since the mean value of $u$ for an Ornstein-Uhlenbeck process decays exponentially, and so the process mean velocity is $\overline{u}=0$.

All the histograms in figure \ref{fig:pdfU01} ($U=0.1\:\mathrm{m/s}$) are reasonably-well fitted by the black solid curves, which represent lognormal distributions. This fit demonstrates the data's heavy 'tails', indicating that the probability for anomalously long lifetime is much larger compared with normal distributions with the same mean and variance. As the air RMS velocity increases (figure \ref{fig:pdfU03}), the lifetime variability is significantly enhanced. Recalling that lognormal distributions appear as Gaussians on a logarithmic scale, we note that the PDFs for $R_{0}=35\:\mathrm{\mu m}$ in figure \ref{fig:pdfU03} deviate strongly from lognormal distribution, displaying substantially fatter tails. To showcase this deviation, we fit these PDFs with log-lognormal distributions, denoted by dashed black curves. 

To understand the remarkable statistics in figure \ref{fig:PDFs}, we separate the discussion on the lifetime of large ($R_{0}=50\:\mathrm{\mu m}$) and small ($R_{0}=20\:\mathrm{\mu m}$) droplets. The large droplets PDFs at varying $\phi$ are grouped together, indicating that vaporization only weakly affects their dynamics. These droplets reach the ground well before the vaporization terminates, and their motion is approximately ballistic regardless of relative humidity. The PDFs in this case are skewed due to the absorbing boundary condition at $z=0$, inducing an asymmetry in the effect of air velocity -- which fluctuates about a zero mean -- on the droplet motion. Indeed, increasing the initial height above the ground allows more time for the air velocity to change direction through the droplet airborne lifetime and entrain it more symmetrically, resulting in convergence towards normal statistics.

Small droplets, on the other hand, are less affected by gravity and do not fall a significant distance, on the average, during their vaporization. The fat tails in their PDFs derive from the nonlinear interplay between vaporization and the drag force acting to entrain droplets to the air flow. At early times, when a droplet vaporizes and $S\left(t\right)$ decreases, the drag force $\left(u-v\right)/C\,S\left(t\right)$ in (\ref{eq:dv ND}) increases non-linearly. The symmetry in $u$ about a zero mean, imposed by the Ornstein-Uhlenbeck process, then leads to an asymmetric effect on the lifetime.

Each of the two effects described above -- finite domain for large droplets and vaporization for small droplets -- leads to a departure from normal distribution. For intermediate size droplets ($R_{0}=35\:\mathrm{\mu m}$), sufficiently large air velocity can trigger a combined effect that dramatically increases the probability for anomalously long lifetimes, as manifested by the $\phi=0.7$ PDF in figure \ref{fig:pdfU03}. The average lifetime for these relatively large droplets is 45 seconds, however 11\% of all droplets are predicted to remain airborne more than 90 seconds. For comparison, less than 1\% of droplets for the equivalent PDF in figure \ref{fig:pdfU01} remain airborne after 90 seconds. As such large droplets can potentially carry a significant viral load, the non-negligible probabilities for anomalously long lifetime can have an effect on virus transmission.

\section{Concluding remarks}

The results throughout show that increasing the relative humidity shortens the airborne lifetime of droplets, which can potentially lower the risk of respiratory transmission. This is indeed the case provided that such increase does not elongate the pathogen viability, which increases the risk \cite{Huynh2022}. The statistical analysis, in which a stochastic process was used to represent indoor turbulence, proposes that even modest intensity of turbulence significantly increases the probability for very long droplet airborne lifetime. This can be a cause for concern if droplets are entrained in ventilation-induced vortices, to be balanced by the clear need for ventilation to discharge droplets with a mean flow of air.

\begin{acknowledgements}
This research was supported by EPSRC Programme Grant EP/R045046/1: Probing Multiscale Complex Multiphase Flows with Positrons for Engineering and Biomedical Applications (PI: Prof. M. Barigou, University of Birmingham).
\end{acknowledgements}

\section*{AUTHOR DECLARATIONS}
\subsection*{Conflict of Interest}
The authors have no conflicts to disclose.

\section*{Data availability}
The data that support the findings of this study are available
from the corresponding author upon reasonable request.

\appendix

\section{Analytic solution for free-falling droplets in quiescent air} \label{ap:A}
We derive an analytic solution for the descent of free-falling droplets by solving the system (\ref{eq:ODE system}) with the following simplifications: setting the air velocity to $u=0$ (quiescent air), and approximating the droplet size evolution $S\left(t\right)$ using the piecewise form (\ref{eq:S piecewise}). The system then simplifies to
\begin{align}
\frac{\d z}{\d t}&=v, \label{eq:dz supp}\\
\frac{\d v}{\d t}&=-\frac{v}{C\,S\left(t\right)}-g, \label{eq:dv supp}
\end{align}
with
\begin{equation}
S\left(t\right)=\begin{cases}
1-\left|\alpha\right| t, & t<t_{eq}\\
S_{eq}, & t>t_{eq}
\end{cases},
\end{equation}
subject to the initial conditions $z\left(t=0\right)=z_{0}$ and $v\left(t=0\right)=v_{0}$. The equations are solved successively, with the solution to (\ref{eq:dv supp}) substituted into (\ref{eq:dz supp}) and integrated, to obtain a closed-form expression. We simplify the expression, which is tedious and is therefore not written explicitly, by employing several asymptotic approximations: based on experimental measurements of saliva droplet vaporization \cite{Lieber2021}, we note that $\left|\alpha\right|\sim10^{-4}-10^{-2}$ and $S_{eq}\sim0.02-0.2$ for the realisable range of air temperature and humidity. Further, for $R_{0}\gtrsim14\,\mathrm{\mu m}$ we have $Kn\lesssim0.05$. Accordingly, we consider the asymptotic limits $\left|\alpha\right|\ll1$, $S_{eq}^{3/2}\ll1$, and $Kn^{2}\ll1$, to finally obtain  
\begin{equation}
z\left(t\right)=\begin{cases}
z_{0}+\left[v_{0}\left(1+\mathcal{A}Kn\right)+g\left(1+2\mathcal{A}Kn\right)\right]\times\\
\left(1-S_{L}^{1/2}\left[\frac{S_{L}^{1/2}+\mathcal{A}Kn}{1+\mathcal{A}Kn}\right]^{2/\left|\alpha\right|}\right)-\\
\frac{g}{6\left|\alpha\right|}\left(3\left[1-S_{L}^{2}\right]+4\mathcal{A}Kn\left[1-S_{L}^{3/2}\right]\right), & t<t_{eq}\\
z_{0}+v_{0}\left(1+\mathcal{A}Kn\right)-\frac{g\left(3+4\mathcal{A}Kn\right)}{6\left|\alpha\right|}-\\
g\left(S_{eq}+\mathcal{A}Kn\sqrt{S_{eq}}\right)\left(t-t_{eq}\right), & t>t_{eq}
\end{cases}
\label{s4}
\end{equation}
where $t_{eq}=\left(1-S_{eq}\right)/\left|\alpha\right|$ and $S_{L}\left(t\right)=1-\left|\alpha\right| t$.

\section{Dependency of $\alpha$ on relative humidity} \label{ap:B}
We begin by rewriting the definition of the $D^2$ vaporization rate $\alpha$,
\begin{equation}
\alpha=-\frac{4\vartheta_{M}\mathcal{F}\Phi\left(\theta_{D2}\right)}{9\lambda}.
\end{equation}
Recalling that $\vartheta_{M},\mathcal{F}$ and $\lambda$ are constants, $\alpha$ varies only according to $\Phi$, which is generally a function of both $\theta$ and $S$ and is given as Eq.\ (\ref{eq:Phi}). However, during the first two vaporization stages the droplet is large enough so that both effects involving $S$ are negligible -- namely surface tension $\eta S^{-1/2}$ and composition-driven increase in boiling temperature $\beta S^{-3/2}$ -- and hence
\begin{align}
\Phi\left(\theta_{D2}\right)&=\exp\left[-\mathcal{H}_{v}\left(\frac{1}{\theta_{D2}+1}-\frac{1}{T_{b}}\right)\right]/\left(\theta_{D2}+1\right)\nonumber \\ &-\phi\exp\left[-\mathcal{H}_{v}\left(1-\frac{1}{T_{b}}\right)\right]. \label{eq:Phi supp}
\end{align}
The temperature $\theta_{D2}\leq0$ reflects a balance between vaporization and convection, acting to decrease and increase the droplet temperature, respectively. In practice, droplets cool to within several degrees below the air temperature, which in scaled form translates to very small variations from zero. Accordingly, we expand (\ref{eq:Phi supp}) as a series near $\theta_{D2}=0$,
\begin{equation}
\Phi\left(\theta_{D2}\right)=\left(1-\phi\right)\exp\left[-\mathcal{H}_{v}\left(1-\frac{1}{T_{b}}\right)\right]+O\left(\theta_{D2}\right),
\end{equation}
showing that $\Phi\left(\theta_{D2}\right)$, to leading order, is linearly dependent on relative humidity. In turn, we find that $\left|\alpha\right|$ decreases nearly linearly with $\phi$.

%\nocite{*}
%\bibliography{DropletDynamics}% Produces the bibliography via BibTeX.

\end{document}